\documentclass[a4paper,nofootinbib,prl]{revtex4}%
\usepackage{amsmath}
\usepackage{amsfonts}
\usepackage{amssymb}
\usepackage{graphicx}%
\begin{document}
\title{Extracting the Maxwell charge from the Wheeler-DeWitt equation}
\author{Remo Garattini}
\email{Remo.Garattini@unibg.it}
\affiliation{Universit\`{a} degli Studi di Bergamo, Facolt\`{a} di Ingegneria,}
\affiliation{Viale Marconi 5, 24044 Dalmine (Bergamo) Italy}
\affiliation{and I.N.F.N. - sezione di Milano, Milan, Italy.}

\begin{abstract}
We consider the Wheeler-De Witt equation as a device for finding eigenvalues
of a Sturm-Liouville problem. In particular, we will focus our attention on
the electric (magnetic) Maxwell charge. In this context, we interpret the
Maxwell charge as an eigenvalue of the Wheeler-De Witt equation generated by
the gravitational field fluctuations. A variational approach with Gaussian
trial wave functionals is used as a method to study the existence of such an
eigenvalue. We restrict the analysis to the graviton sector of the
perturbation. We approximate the equation to one loop in a Schwarzschild
background and a zeta function regularization is involved to handle with
divergences. The regularization is closely related to the subtraction
procedure appearing in the computation of Casimir energy in a curved
background. A renormalization procedure is introduced to remove the infinities
together with a renormalization group equation.

\end{abstract}
\maketitle

\section{Introduction}

In 1955, John Archibald Wheeler\cite{Wheeler} considered the possibility that
the gravitational coupled to the electromagnetic field could lead to a
sourceless solution termed \textquotedblleft\textit{geon}\textquotedblright.
Further studies in this direction gave birth to particular ideas such as
\textquotedblleft\textit{mass without mass}\textquotedblright\ and
\textquotedblleft\textit{charge without charge}\textquotedblright, where
fluctuations of the gravitational field were thought as responsible of the
generation of elementary particles. It is clear that, if such a possibility
exists, this is encoded in the Einstein's field equations. These equations are
simply summarized by
\begin{equation}
R_{\mu\nu}-\frac{1}{2}g_{\mu\nu}R+\Lambda_{c}g_{\mu\nu}=\kappa T_{\mu\nu},
\label{ein}%
\end{equation}
where $T_{\mu\nu}$ is the energy-momentum tensor of some matter fields,
$\kappa=8\pi G$ with $G$ the Newton's constant and $\Lambda_{c}$ is the
cosmological constant. The idea is to recognize the gravitational field as a
fundamental field and see what implications we have on the cosmological
constant and on the matter fields. In Ref.\cite{Remo}, we have applied this
concept to the cosmological constant. In particular, we have considered the
cosmological constant as an eigenvalue of an associated Sturm-Liouville
problem, even in presence of a massive graviton\cite{Remo1}. Motivated by this
result, in this paper we would like to apply the same approach to the Maxwell
charge. To do this, we need to introduce the Wheeler-DeWitt equation
(WDW)\cite{De Witt}. The WDW equation can be extracted from the Einstein's
field equations with and without matter fields in a very simple way. If we
introduce a time-like unit vector $u^{\mu}$ such that $u\cdot u=-1$, then
after a little rearrangement of Eqs.$\left(  \ref{ein}\right)  $, we get:%
\begin{equation}
\mathcal{H}_{0}\mathcal{=}\left(  2\kappa\right)  G_{ijkl}\pi^{ij}\pi
^{kl}-\frac{1}{2\kappa}\!{}\!\,\sqrt{g}\,^{3}R=0, \label{HLambda}%
\end{equation}
for the sourceless case and in absence of a cosmological term.%
\begin{equation}
\mathcal{H}_{\Lambda}\mathcal{=}\left(  2\kappa\right)  G_{ijkl}\pi^{ij}%
\pi^{kl}-\frac{\sqrt{g}}{2\kappa}\!{}\!\left(  \,^{3}R-2\Lambda_{c}\right)
=0, \label{HL}%
\end{equation}
for the sourceless case and in presence of a cosmological term.%
\begin{equation}
\mathcal{H}_{Q}\mathcal{=}\left(  2\kappa\right)  G_{ijkl}\pi^{ij}\pi
^{kl}-\frac{\sqrt{g}}{2\kappa}\!{}\!\left(  \,^{3}R-\mathcal{H}_{M}\right)
=0, \label{HQ}%
\end{equation}
with a matter term and in absence of a cosmological constant. Note the formal
similarity between Eqs.$\left(  \ref{HL}\right)  $ and $\left(  \ref{HQ}%
\right)  $. $G_{ijkl}$ is the \textit{supermetric} defined as%
\begin{equation}
G_{ijkl}=\frac{1}{2}(g_{ik}g_{jl}+g_{il}g_{jk}-g_{ij}g_{kl})
\end{equation}
and $^{3}R$ is the scalar curvature in three dimensions. $\pi^{ij}$ is called
the supermomentum. This is the time-time component of Eqs.$\left(
\ref{ein}\right)  $. It represents the invariance under \textit{time}
reparametrization and it works as a constraint at the classical level. On the
other hand, the form of $\mathcal{H}_{M}$ for the electromagnetic field can be
obtained with the same method used for the pure gravitational field and it can
be written as
\begin{equation}
\mathcal{H}_{M}=\kappa\sqrt{^{3}g}T_{\alpha\beta}u^{\alpha}u^{\beta},
\end{equation}
where
\begin{equation}
T_{\alpha\beta}=\frac{1}{4\pi}\left[  F_{\alpha\gamma}F_{\beta}^{\gamma}%
-\frac{1}{4}g_{\alpha\beta}F_{\gamma\delta}F^{\gamma\delta}\right]
\end{equation}
and $F_{\alpha\gamma}=\partial_{\alpha}A_{\gamma}-\partial_{\gamma}A_{\alpha}%
$. $A_{\alpha}$ is the electromagnetic potential which, in the case of a pure
electric field assumes the form $A_{\alpha}=\left(  Q_{e}/r,0,0,0\right)  $
while in the case of pure magnetic field, the form is $A_{\alpha}=\left(
0,-Q_{m}\sin\theta,0,0\right)  $. $Q_{e}$ and $Q_{m}$ are the electric and
magnetic charge respectively. Both of them contribute in the same way to the
gravitational potential of Eq. $\left(  \ref{p12}\right)  $. For the electric
charge, the on-shell contribution of $T_{\alpha\beta}u^{\alpha}u^{\beta}$ is
\begin{equation}
T_{\alpha\beta}u^{\alpha}u^{\beta}=\frac{1}{8\pi}\left(  F_{01}\right)
^{2}=\frac{1}{8\pi}\frac{Q_{e}^{2}}{r^{4}}=\rho_{e}, \label{elec}%
\end{equation}
while when we consider the magnetic charge, we get%
\begin{equation}
T_{\alpha\beta}u^{\alpha}u^{\beta}=\frac{1}{8\pi}\left(  F_{23}\right)
^{2}=\frac{1}{8\pi}\frac{Q_{m}^{2}}{r^{4}}=\rho_{m}. \label{magn}%
\end{equation}
Thus, the classical constraint $\mathcal{H}_{Q}$ for a Maxwell charge becomes%
\begin{equation}
\mathcal{H}_{Q}\mathcal{=}\left(  2\kappa\right)  G_{ijkl}\pi^{ij}\pi
^{kl}-\frac{\sqrt{g}}{2\kappa}\!{}\!\left(  \,^{3}R-\kappa\sqrt{^{3}%
g}T_{\alpha\beta}u^{\alpha}u^{\beta}\right)  =0.
\end{equation}
If $\mathcal{H}_{Q}$ is promoted to an operator, then the following quantum
constraint%
\begin{equation}
\mathcal{H}_{Q}\Psi\mathcal{=}0 \label{WDW}%
\end{equation}
is imposed. This is known as the WDW equation. The WDW can be rearranged in
such a way to show a more useful aspect. Indeed, if we integrate Eq.$\left(
\ref{WDW}\right)  $ over the hypersurface $\Sigma$ obtained with the help of
the Arnowitt-Deser-Misner variables (ADM)\cite{ADM} and we define%
\begin{equation}
\hat{Q}_{\Sigma}=\left(  2\kappa\right)  G_{ijkl}\pi^{ij}\pi^{kl}-\frac
{\sqrt{g}}{2\kappa}\!{}\,^{3}R,
\end{equation}
then Eq.$\left(  \ref{WDW}\right)  $ can be cast into the following form%
\begin{equation}
\frac{\left\langle \Psi\left\vert \int_{\Sigma}d^{3}x\hat{Q}_{\Sigma
}\right\vert \Psi\right\rangle }{\left\langle \Psi|\Psi\right\rangle }%
=-\frac{\left\langle \Psi\left\vert \int_{\Sigma}d^{3}x\left(  \,\sqrt{^{3}%
g}T_{\alpha\beta}u^{\alpha}u^{\beta}\right)  \right\vert \Psi\right\rangle
}{2\left\langle \Psi|\Psi\right\rangle }. \label{WDW0}%
\end{equation}
Eq.$\left(  \ref{WDW0}\right)  $ has been obtained by multiplying Eq.$\left(
\ref{WDW}\right)  $ by $\Psi^{\ast}\left[  g_{ij}\right]  $ and by
functionally integrating over the three spatial metric $g_{ij}$. To fix the
ideas, let us consider the electric case. Thus, if we substitute the
expression $\left(  \ref{elec}\right)  $ into Eq.$\left(  \ref{WDW0}\right)
$, we get%
\begin{equation}
\frac{\left\langle \Psi\left\vert \int_{\Sigma}d^{3}x\hat{Q}_{\Sigma
}\right\vert \Psi\right\rangle }{\left\langle \Psi|\Psi\right\rangle }%
=-\frac{1}{2}\int_{\Sigma}d^{3}x\sqrt{^{3}g}\rho_{e}. \label{Hel}%
\end{equation}
It is immediately clear that a classical solution is represented by the
Reissner-Nordstr\"{o}m (RN) metric, whose form is%
\begin{equation}
ds^{2}=-f\left(  r\right)  dt^{2}+\frac{dr^{2}}{f\left(  r\right)  }%
+r^{2}\left(  d\theta^{2}+\sin^{2}\theta d\phi^{2}\right)  ,
\end{equation}
where the gravitational potential is expressed by%
\begin{equation}
f\left(  r\right)  =1-\frac{2MG}{r}+\frac{G\left(  Q_{e}^{2}+Q_{m}^{2}\right)
}{r^{2}}. \label{p12}%
\end{equation}
Nevertheless, we are not interested in finding corrections to the RN metric,
rather we want to use Eq.$\left(  \ref{Hel}\right)  $ as a device to find
consistent solutions of an electric/magnetic charge generated by quantum
fluctuations of the gravitational field. This approach is not completely new,
it appears naturally when black hole mass quantization is discussed in some
specific models even including a charge\cite{BDK,VazWit,MR}. Nevertheless, in
this approach we will never deal with RN black holes. On the other hand, it
appears that our approach seems to be closer to Ref.\cite{Sones}, where a
quantum analysis of Wheeler's geons\cite{Wheeler} was carried out, even if a
real quantum computation seems to be absent. The semi-classical procedure
followed in this work relies heavily on the formalism outlined in
Refs.\cite{Remo,Remo1}. The computation was realized through a variational
approach with Gaussian trial wave functionals. A zeta function regularization
is used to deal with the divergences, and a renormalization procedure is
introduced, where the finite one loop is considered as a self-consistent
source for traversable wormholes. Rather than reproduce the formalism, we
shall refer the reader to Refs.\cite{Remo,Remo1} for details, when necessary.

\section{One loop Charge evaluation}

We can write the background metric in the following form%
\begin{equation}
ds^{2}=-N^{2}\left(  r\right)  dt^{2}+\frac{dr^{2}}{1-\frac{b\left(  r\right)
}{r}}+r^{2}\left(  d\theta^{2}+\sin^{2}\theta d\phi^{2}\right)  ,
\label{metric}%
\end{equation}
where $r\in\left[  r_{t},+\infty\right)  $ and%
\begin{equation}
b\left(  r_{t}\right)  =r_{t}.
\end{equation}
$r_{t}$ is termed the throat. $N\left(  r\right)  $ is the \textquotedblleft%
\textit{lapse function}\textquotedblright\ playing the role of the
\textquotedblleft\textit{redshift function}\textquotedblright, while $b\left(
r\right)  $ is termed \textquotedblleft\textit{shape function}%
\textquotedblright. We take into account the total regularized one loop energy
given by
\begin{equation}
E^{TT}=2\int_{r_{0}}^{\infty}\,dr\frac{r^{2}}{\sqrt{1-b(r)/r}}\,\left[
\rho_{1}(\varepsilon)+\rho_{2}(\varepsilon)\right]  \,.
\end{equation}
The energy densities, $\rho_{i}(\varepsilon)$ (with $i=1,2$), are defined as%
\[
\rho_{i}(\varepsilon)=\frac{1}{4\pi}\mu^{2\varepsilon}\int_{\sqrt{U_{i}(r)}%
}^{\infty}\,d\tilde{E}_{i}\,\frac{\tilde{E}_{i}^{2}}{\left[  \tilde{E}_{i}%
^{2}-m_{i}^{2}(r)\right]  ^{\varepsilon-1/2}}%
\]%
\begin{equation}
=-\frac{m_{i}^{4}(r)}{64\pi^{2}}\left[  \frac{1}{\varepsilon}+\ln\left(
\frac{\mu^{2}}{m_{i}^{2}\left(  r\right)  }\right)  +2\ln2-\frac{1}{2}\right]
. \label{energy}%
\end{equation}
where we have defined two r-dependent effective masses $m_{1}^{2}\left(
r\right)  $ and $m_{2}^{2}\left(  r\right)  $, which can be cast in the
following form%
\begin{equation}
\left\{
\begin{array}
[c]{c}%
m_{1}^{2}\left(  r\right)  =m_{L}^{2}\left(  r\right)  +m_{1,S}^{2}\left(
r\right) \\
\\
m_{2}^{2}\left(  r\right)  =m_{L}^{2}\left(  r\right)  +m_{2,S}^{2}\left(
r\right)
\end{array}
\right.  ,
\end{equation}
where%
\begin{equation}
m_{L}^{2}\left(  r\right)  =\frac{6}{r^{2}}\left(  1-\frac{b\left(  r\right)
}{r}\right)  \label{mL}%
\end{equation}
and%
\begin{equation}
\left\{
\begin{array}
[c]{c}%
m_{1,S}^{2}\left(  r\right)  =\left[  \frac{3}{2r^{2}}b^{\prime}\left(
r\right)  -\frac{3}{2r^{3}}b\left(  r\right)  \right] \\
m_{2,S}^{2}\left(  r\right)  =\left[  \frac{1}{2r^{2}}b^{\prime}\left(
r\right)  +\frac{3}{2r^{3}}b\left(  r\right)  \right]
\end{array}
\right.  . \label{m12s}%
\end{equation}
We refer the reader to Refs. \cite{Remo,Remo1} for the deduction of these
expressions in the Schwarzschild case. The zeta function regularization method
has been used to determine the energy densities, $\rho_{i}$. It is interesting
to note that this method is identical to the subtraction procedure of the
Casimir energy computation, where the zero point energy in different
backgrounds with the same asymptotic properties is involved. In this context,
the additional mass parameter $\mu$ has been introduced to restore the correct
dimension for the regularized quantities. Note that this arbitrary mass scale
appears in any regularization scheme. Eq.$\left(  \ref{Hel}\right)  $ for the
energy density becomes%
\begin{equation}
\rho_{e}=\rho_{1}(\varepsilon)+\rho_{2}(\varepsilon)\,. \label{rho}%
\end{equation}
Taking into account Eq.$\left(  \ref{energy}\right)  $, then Eq.$\left(
\ref{rho}\right)  $ yields the following relationship%
\[
\rho_{e}=\frac{1}{8\pi}\frac{Q_{e}^{2}}{r^{4}}%
\]%
\begin{equation}
\hspace{-1.5cm}=\frac{1}{64\pi^{2}}\left[  \frac{1}{\varepsilon}\,\left[
m_{1}^{4}(r)+m_{2}^{4}(r)\right]  +m_{1}^{4}(r)\ln\left(  \left\vert
\frac{4\mu^{2}}{m_{1}^{2}(r)\sqrt{e}}\right\vert \right)  +m_{2}^{4}%
(r)\ln\left(  \left\vert \frac{4\mu^{2}}{m_{2}^{2}(r)\sqrt{e}}\right\vert
\right)  \right]  . \label{charge}%
\end{equation}
It is essential to renormalize the divergent energy by absorbing the
singularity in the classical quantity, by redefining the bare classical charge
$Q_{e}^{2}$ as
\begin{equation}
Q_{e}^{2}\rightarrow Q_{e,0}^{2}+\frac{\left[  m_{1}^{4}(r)+m_{2}%
^{4}(r)\right]  }{32\varepsilon\pi^{2}}r^{4}.
\end{equation}
Using this, Eq. $\left(  \ref{charge}\right)  $ takes the form
\begin{equation}
Q_{e,0}^{2}\left(  \mu\right)  =\frac{r^{4}}{32\pi}\left[  m_{1}^{4}\left(
r\right)  \ln\left(  \left\vert \frac{4\mu^{2}}{m_{1}^{2}\left(  r\right)
\sqrt{e}}\right\vert \right)  +m_{2}^{4}\left(  r\right)  \ln\left(
\left\vert \frac{4\mu^{2}}{m_{1}^{2}\left(  r\right)  \sqrt{e}}\right\vert
\right)  \right]  =\left(  Q_{eff}^{2}\right)  ^{TT}\left(  \mu,r\right)  .
\label{charge1}%
\end{equation}

To avoid the dependence on the arbitrary mass scale $\mu$ in Eq.$\left(
\ref{charge1}\right)  $, we adopt the renormalization group equation and we
impose that\cite{RGeq}%
\begin{equation}
\mu\frac{\partial}{\partial\mu}Q_{e,0}^{2}\left(  \mu\right)  =\mu\frac
{d}{d\mu}\left(  Q_{eff}^{2}\right)  ^{TT}\left(  \mu,r\right)  .
\end{equation}
Solving it we find that the renormalized squared charge $Q_{e,0}^{2}\left(
\mu\right)  $ should be treated as a running one in the sense that it varies
provided that the scale $\mu$ is changing
\begin{equation}
Q_{e,0}^{2}\left(  \mu,r\right)  =Q_{e,0}^{2}\left(  \mu_{0},r\right)
+\frac{r^{4}}{16\pi^{2}}\left(  m_{1}^{4}(r)+m_{2}^{4}(r)\right)  \ln\frac
{\mu}{\mu_{0}}. \label{RGsol}%
\end{equation}
Substituting Eq.$\left(  \ref{RGsol}\right)  $ into Eq.$\left(  \ref{charge1}%
\right)  $ we find%
\begin{equation}
Q_{e,0}^{2}\left(  \mu_{0},r\right)  =-\frac{r^{4}}{32\pi}\left[  m_{1}%
^{4}\left(  r\right)  \ln\left(  \frac{m_{1}^{2}(r)\sqrt{e}}{4\mu_{0}^{2}%
}\right)  +m_{2}^{4}\left(  r\right)  \ln\left(  \frac{m_{2}^{2}(r)\sqrt{e}%
}{4\mu_{0}^{2}}\right)  \right]  . \label{charge2}%
\end{equation}
Using Eqs.$\left(  \ref{mL},\ref{m12s}\right)  $, Eq.$\left(  \ref{charge2}%
\right)  $ becomes%
\[
Q_{e,0}^{2}\left(  \mu_{0},r\right)  =-\frac{r^{4}}{32\pi}\left\{  \left(
m_{L}^{2}\left(  r\right)  +m_{1,S}^{2}\left(  r\right)  \right)  ^{2}\left[
\ln\left(  \frac{m_{L}^{2}\left(  r\right)  +m_{1,S}^{2}\left(  r\right)
}{4\mu_{0}^{2}}\sqrt{e}\right)  \right]  \right.
\]%
\begin{equation}
\left.  +\left(  m_{L}^{2}\left(  r\right)  +m_{2,S}^{2}\left(  r\right)
\right)  ^{2}\left[  \ln\left(  \frac{m_{L}^{2}\left(  r\right)  +m_{2,S}%
^{2}\left(  r\right)  }{4\mu_{0}^{2}}\sqrt{e}\right)  \right]  \right\}  .
\label{charge3}%
\end{equation}
Even if this result is valid for an arbitrary function $b\left(  r\right)  $,
we fix our attention to the Schwarzschild metric, motivated by the fact that
this is the most simple spherically symmetric solution of the Einstein's field
equations \textit{without a source term}. In this case, we get $b\left(
r\right)  =2MG=r_{t}$. It is straightforward to see that for large distances,
$m_{L}^{2}\left(  r\right)  $ dominates and $Q_{e,0}^{2}\left(  \mu
_{0},r\right)  $ diverges. So, the validity of this result is confined to stay
close to the throat. When we approach $r_{t}$, $m_{L}^{2}\left(  r\right)
\rightarrow0$ and $m_{1,S}^{2}\left(  r\right)  <0$. Precisely, $m_{1,S}%
^{2}\left(  r\right)  $ becomes negative when $r\in\left[  r_{t},\frac{5}%
{4}r_{t}\right]  $ and $m_{1,S}^{2}\left(  r\right)  =-m_{2,S}^{2}\left(
r\right)  =-3r_{t}/2r^{3}$. In such a range, Eq.$\left(  \ref{charge3}\right)
$ simplifies to%
\begin{equation}
Q_{e,0}^{2}\left(  \mu_{0},r\right)  =-\frac{r^{4}}{16\pi}\left(  \frac
{3r_{t}}{2r^{3}}\right)  ^{2}\ln\left(  \frac{3r_{t}\sqrt{e}}{8r^{3}\mu
_{0}^{2}}\right)  .
\end{equation}
In order to remove the dependence on $r$, we compute%
\begin{equation}
\frac{\partial Q_{e,0}^{2}\left(  \mu_{0},r\right)  }{\partial r}%
=0\qquad\Longrightarrow\qquad\frac{3r_{t}e}{8\mu_{0}^{2}}=\bar{r}^{3}%
\end{equation}
and%
\begin{equation}
Q_{e,0}^{2}\left(  \mu_{0},\bar{r}\right)  =\frac{\bar{r}^{4}\mu_{0}^{4}}{2\pi
e^{2}}=\left(  \frac{3r_{t}\mu_{0}}{8\sqrt{e}}\right)  ^{\frac{4}{3}}\frac
{1}{2\pi}. \label{charge4}%
\end{equation}
Since%
\begin{equation}
r\in\left[  r_{t},\frac{5}{4}r_{t}\right]  \qquad\Longrightarrow\qquad
\sqrt{\frac{3e}{8r_{t}^{2}}}\geq\mu_{0}\geq\sqrt{\frac{24e}{125r_{t}^{2}}},
\end{equation}
we find the following bound%
\begin{equation}
2.\,\allowbreak2\times10^{-2}=\frac{9}{128\pi}\geq Q_{e,0}^{2}\left(  \mu
_{0},\bar{r}\right)  \geq\frac{9}{200\pi}=1.\,\allowbreak4\times10^{-2}.
\label{bound}%
\end{equation}
Note that the fine structure constant is $\frac{1}{137}=\allowbreak
.7\,\allowbreak3\times10^{-2}$\footnote{See also Ref.\cite{Rosales} for
another approach based on the WDW equation.}. A comment to this result is in
order. From Eq.$\left(  \ref{charge4}\right)  $ and the bound $\left(
\ref{bound}\right)  $, we infer that once the charge has been created form
quantum fluctuations, it never disappears, unless the throat is large as the
whole universe. Secondly, the bound is very close to the fine structure
constant, but it has not the desired value. This is a good news, because the
computation is limited to the graviton and even if it is represented by TT
modes which are gauge invariant, they do not represent the whole perturbation,
only the leading one. Therefore, it is likely that the new input will correct
the value of the bound. In particular, it is expected that trace modes could
screen the Maxwell charge because of the opposite sign of the spin 0 term.

\section{Conclusions}

In this letter, we have considered the possibility that an electric/magnetic
charge be generated by quantum fluctuations of the pure gravitational field.
The calculational kit is based on a variational version of the Sturm-Liouville
problem, already applied in the cosmological context\cite{Remo,Remo1}. In this
context, as in other contexts examined with this approach, e.g.,
self-sustained wormholes\cite{Remo2}, we have put the accent on the
fluctuations of the gravitational field which act as a source for the matter
fields. This still is in the spirit of the Einstein's field equations and the
idea of unification of the forces, but it is in contrast with the Sakharov's
induced gravity\cite{Sakharov}. In such a theory, the low-energy effective
action $\Gamma\left[  g\right]  $ is defined as a quantum average of the
constituent matter fields $\Phi$ propagating in a given gravitational
background $g$%
\begin{equation}
\exp\left(  -\Gamma\left[  g\right]  \right)  =\int\mathcal{D}\Phi\exp\left(
-S\left[  \Phi,g\right]  \right)  .
\end{equation}
The Sakharov's basic assumption is that the gravitational field becomes
dynamical only as the result of quantum effects of the constituents fields.
This theory has the pleasant feature of being renormalizable. However, its
application is rather limited to some specific problems like black hole
entropy\cite{FrolovFursaev}. It is interesting to note that our approach
provides also a nonvanishing magnetic charge. This result could support the
idea that at the Planck scale magnetic monopoles could exist. Nevertheless, if
this is true, it is straightforward to see that a suppression mechanism at low
energy should emerge. Moreover, we have to observe that the choice of the
Gaussian wave functional corresponds to a \textquotedblleft\textit{ground
state}\textquotedblright\ functional and consistently we obtain only one
eigenvalue. Different choices of the wave functional correspond to different
boundary conditions. Moreover, different forms of the Gaussian wave
functionals can be considered to form \textquotedblleft\textit{excited
states}\textquotedblright. Finally, we want to remark that the Schwarzschild
choice has been made because this is the simplest spherically symmetric
solution of the Einstein's field equations without matter fields, which is
exactly what we need to generate a charge from gravity.

\section{Acknowledgments}

I wish to thank S. Capozziello for useful comments and discussions.

\end{document}